\documentclass{iopart}

\usepackage{iopams}
\usepackage{graphicx}
\global\arraycolsep=2pt

\begin{document}

\title[How does Casimir energy fall? III.] 
{How does Casimir energy fall? III. Inertial forces on vacuum energy.} 
\author{K V Shajesh, Kimball A Milton, Prachi Parashar and Jeffrey A Wagner}
\address{Oklahoma Center for High Energy Physics and
Homer L. Dodge Department of Physics and Astronomy, University of Oklahoma,
Norman, OK 73019-2061 USA.}
\ead{shajesh@nhn.ou.edu \hspace{3cm} Date: \today}

\begin{abstract}
We have recently demonstrated that Casimir energy due to parallel plates,
including its divergent parts, falls like conventional mass
in a weak gravitational field.
The divergent parts were suitably interpreted as renormalizing
the bare masses of the plates.
Here we corroborate our result regarding the inertial nature of 
Casimir energy by calculating the centripetal force on a Casimir 
apparatus rotating with constant angular speed. 
We show that the centripetal force is independent of 
the orientation of the Casimir apparatus
in a frame whose origin is at the center of inertia
of the apparatus.
\end{abstract}

%\tableofcontents

%\pacs{03.70.+k, 04.20.Cv, 04.25.Nx, 03.30.+p} 

%-------------------------------------------
\section{Introduction}

Two uncharged parallel conducting plates attract.
Casimir in 1948~\cite{Casimir:1948dh} demonstrated that 
this force of attraction can be attributed to the
quantum vacuum energy associated with the electromagnetic field
between the conducting plates. Precision measurement of 
this force since 1997~\cite{Lamoreaux:1996wh}
being invoked as an evidence of zero-point energy of the vacuum
has been questioned in~\cite{Jaffe:2005vp}.
The controversy arises due to the presence of divergences
which make it difficult to extract self energies for 
single bodies.
The local energy density, a component of the stress tensor,
serves as a source to gravity. Surface divergences in the
local energy densities cause a serious difficulty in attempts
to understand how the vacuum energy interacts with gravity. 

In Fall I~\cite{Fulling:2007xa} 
we demonstrated, using variational principles,
that the gravitational force on Casimir energy is the same 
as that on a conventional mass.
In Fall II~\cite{Milton:2007ar}
we considered the motion of a Casimir apparatus 
undergoing a constant acceleration.
The Casimir apparatus consisted of semi-transparent parallel plates, 
realized by background fields consisting of delta function potentials,
which induced boundary conditions on a massless scalar field.
The motion of the apparatus was studied in the 
accelerated frame of reference, described by Rindler coordinates,
in which it
experiences a pseudo-force. This fictitious force is 
equated to the gravitational force by the principle of equivalence.
We concluded that Casimir energy, including the divergent 
contributions, falls under gravity like conventional mass.
We made the remarkable observation that
the divergent contributions to Casimir energy could be 
suitably interpreted as renormalizing the bare masses of the 
individual Casimir plates.
Saharian et al.~\cite{Saharian:2003fd},
using zeta function technique for regularization,
earlier reached at a similar conclusion
for the finite part of the Casimir energy.
Thus, in Fall I and II, which has been summarized and submitted 
to this proceedings by K A Milton~\cite{Milton:2007hd}, 
we have convincingly demonstrated that Casimir energy falls like 
conventional mass in a weak gravitational field.

To further verify our conclusion we here study 
the centripetal force acting on a rotating Casimir apparatus.
Firstly, we use this opportunity to refine our definition of
the force on the energy associated with the field.
Next, we consider a Casimir apparatus rotating with constant angular
speed such that the surfaces of the plates remain parallel
to the tangent to the circle of rotation.
We use appropriate curvilinear coordinates to transform 
into the accelerated frame of reference. We demonstrate that,
exactly as in the case of Rindler acceleration, 
the Casimir energy, including its divergent parts,
experiences a centripetal force exactly like a conventional mass. 
We express the centripetal acceleration in terms of the 
center of inertia of the apparatus.
Finally, we consider a rotating Casimir apparatus such that the
plates makes an arbitrary angle with respect to the 
tangent to the circle. We demonstrate that the centripetal force 
is independent of the orientation of the Casimir apparatus 
in a frame in which the center of inertia evaluates to zero.

%-------------------------------------------
\section{Force in the local Lorentz coordinates}

Consider the curvilinear coordinates $x^\mu \equiv (t,x,y,z)$ with
the associated metric $g_{\mu\nu}(x)$.
A particular set of coordinates 
$\bar{x}^a \equiv (\bar{t},\bar{x},\bar{y},\bar{z})$
with metric $\eta_{ab} = (-1,+1,+1,+1)$
are called the local Lorentz coordinates.
Let us consider a transformation $\bar{x}^a=\bar{x}^a(x)$
which can always be defined in a local neighbourhood.
Any spatial point ${\bf x}$ in the curvilinear coordinates
is called a point of reference~\cite{moller}.
As a warm-up, and to acquaint ourselves with the notations,
let us define the velocity of the point of reference ${\bf x}$
as measured in the local Lorentz coordinates at time $\bar{t}$. 
To this regard we can write
\begin{eqnarray}
\Delta {\bf \bar{x}} 
&= \Delta {x^i} \frac{\partial {\bf \bar{x}}}{\partial x^i} 
+ \Delta t \frac{\partial {\bf \bar{x}}}{\partial t}, 
\qquad \qquad \Delta \bar{t}
&= \Delta {x^i} \frac{\partial \bar{t}}{\partial x^i}
+ \Delta t \frac{\partial \bar{t}}{\partial t},
\end{eqnarray}
in terms of which the velocity of the point of reference ${\bf x}$
as measured in the local Lorentz coordinates at time $\bar{t}$
will be
\begin{equation}
{\bf v}({\bf \bar{x}}, \bar{t}) 
= \left. \frac{\Delta {\bf \bar{x}}}{\Delta \bar{t}} \right|_{\bf x}
= \left[ \frac{\partial \bar{t}(x)}{\partial t} \right]^{-1}
   \frac{\partial {\bf \bar{x}}(x)}{\partial t},
\label{def-vel}
\end{equation}
because $\Delta {x^i} =0$ for the point of reference.

A specific field, which is an analog of a fluid, is described by 
the corresponding energy-momentum tensor $t_{\mu\nu}(x)$.
The momentum of a small volume containing the field at 
a point of reference ${\bf x}$ as measured in the local 
Lorentz coordinates at time $\bar{t}$ is %will be
\begin{equation}
\bar{P}^a(\bar{x}) 
= d\Sigma_b (\bar{x}) \, t^{ab} (\bar{x})
= d\Sigma_\nu (x) \, t^{\mu\nu} (x) \, 
  \frac{\partial \bar{x}^a(x)}{\partial x^\mu}
= P^\mu(x) \frac{\partial \bar{x}^a(x)}{\partial x^\mu},
\end{equation}
where we have used tensor transformation properties.
The spatial volume elements are constructed as the antisymmetric 
product of three space like vectors,
$d\Sigma_a(\bar{x}) = \epsilon_{abcd} 
   \,\delta_1 \bar{x}^b \,\delta_2 \bar{x}^c \, \delta_3 \bar{x}^d$,
and
$d\Sigma_\mu(x)=\sqrt{- g(x)} \,\epsilon_{\mu\nu\alpha\beta} 
   \,\delta_1 x^\nu \,\delta_2 x^\alpha \, \delta_3 x^\beta$,
where $g(x) = \mbox{det} \,g_{\mu\nu}(x)$.
The change in momentum when the coordinates are displaced will be
\begin{equation}
\Delta \bar{P}^a(\bar{x})
= \Delta \bar{x}^i \frac{\partial}{\partial \bar{x}^i} \bar{P}^a(\bar{x})
+ \Delta \bar{t} \frac{\partial}{\partial \bar{t}} \bar{P}^a(\bar{x})
\end{equation}
and the force density as measured in the local Lorentz
coordinates at time $\bar{t}$ is 
\begin{eqnarray}
\bar{F}^a(\bar{x}) = 
\left. \frac{\Delta \bar{P}^a}{\Delta \bar{t}} \right|_{\bf x}
&=& 
\left. \frac{\Delta \bar{x}^i}{\Delta \bar{t}} \right|_{\bf x}
\left. \frac{\partial}{\partial \bar{x}^i} \bar{P}^a(\bar{x}) \right|_{\bf x}
+ \left. \frac{\partial}{\partial \bar{t}} \bar{P}^a(\bar{x}) \right|_{\bf x}
\nonumber \\
%&=& \left[ \frac{\partial \bar{t}(x)}{\partial t} \right]^{-1}
%\left[ \frac{\partial \bar{x}^i}{\partial t} 
%       \frac{\partial}{\partial \bar{x}^i} \bar{P}^a(\bar{x})
%       + \frac{\partial \bar{t}}{\partial t}
%         \frac{\partial}{\partial \bar{t}} \bar{P}^a(\bar{x}) \right]_{\bf x}
%\nonumber \\
&=& \left[ \frac{\partial \bar{t}(x)}{\partial t} \right]^{-1}
\left[ \frac{\partial}{\partial t} \bar{P}^a(\bar{x}) \right]_{\bf x}
\nonumber \\
&=& \left[ \frac{\partial \bar{t}(x)}{\partial t} \right]^{-1}
    \frac{\partial}{\partial t} 
    \left[ P^\mu(x) \frac{\partial \bar{x}^a(x)}{\partial x^\mu} \right],
\label{def-force}
\end{eqnarray}
where we used \eref{def-vel} in going from first line to the 
second line.

The total force on the energy associated with 
the field as measured in the local Lorentz coordinates
at time $\bar{t}$ is obtained by integrating the force 
density over a surface $S$ described by constant $\bar{t}$. Thus
\begin{eqnarray}
\bar{F}^a(\bar{t}) 
&=& \int_S \left[ \frac{\partial \bar{t}(x)}{\partial t} \right]^{-1}
    \frac{\partial}{\partial t}
    \left[ P^\mu(x) \frac{\partial \bar{x}^a(x)}{\partial x^\mu} \right]
\nonumber \\
&=& \int d^3x 
\left[ \frac{\partial \bar{t}(x)}{\partial t} \right]^{-1}
\,\frac{\partial}{\partial t}
\left[ \sqrt{-g(x)} \,t^{\mu 0}(x) 
       \,\frac{\partial\bar{x}^a(x)}{\partial x^\mu}\right].
\label{FA}
\end{eqnarray}
This probably is rigorously true only for situations when the field 
under consideration is sufficiently localized and thus requires scrutiny.
In the special case when both the metric tensor and the 
energy momentum tensor are independent of the coordinate 
time $t$ the above expression takes the simpler form
\begin{eqnarray}
\bar{F}^a(\bar{t})
&=& \int d^3x \sqrt{-g(x)} \,t^{\mu 0}(x)        
\, \left[ \frac{\partial}{\partial t}
\,\frac{\partial\bar{x}^a(x)}{\partial x^\mu} \right]
\left[ \frac{\partial \bar{t}(x)}{\partial t} \right]^{-1}.
\label{FA-ct}
\end{eqnarray}

%-------------------------------------------
\subsection{Rindler metric}
As an illustrative example let us reconsider the situation 
in Fall II using the above definition. The local Lorentz 
coordinates $\bar{x}^a$ are related to the Rindler coordinates
$x^\mu \equiv (\tau,x,y,\xi)$ by the transformations:
$\bar{t}=\xi\sinh\tau,\;\bar{x}=x,\;\bar{y}=y,\;\bar{z}=\xi\cosh\tau$.
The Rindler metric is 
$g_{\mu\nu}(x) = (-\xi^2,+1,+1,+1)$ which returns 
$g(x) = - \xi^2$. Further for the Casimir apparatus considered 
in Fall II we have only the diagonal components of $t_{\mu\nu}(x)$
to be nonzero which are independent of $\tau$. 
Using these in \eref{FA-ct} we determine the force 
on the vacuum energy associated with the Casimir apparatus to be
\begin{equation}
\frac{F^3}{A}
= \int_0^\infty \frac{d\xi}{\xi} \,\xi^2 \,t^{00}(\xi)  
= \int_0^\infty \frac{d\xi}{\xi} \,\frac{1}{\xi^2} \,t_{00}(\xi)  
= - \int_0^\infty \frac{d\xi}{\xi} \,{t^0}_0(\xi), 
\label{F-rind}
\end{equation}
which leads to the result in Fall II
\begin{equation}
F^3 = \left[E_{\rm a}^{(0)} + E_{\rm b}^{(0)} + E_{\rm cas}^{(0)} \right]g,
\end{equation}
where the first two terms are the divergent energies associated 
with the individual plates and $E_{\rm cas}^{(0)}$ is the Casimir
energy. Further, using the stress tensor for a 
conventional mass,
\begin{equation}
t^{\mu\nu}(x) = m_{\rm a} \frac{1}{\sqrt{-g(x)}} 
\int_{-\infty}^{+\infty} ds \,\delta^{(4)}(x-x_{\rm a}(s))
\frac{dx^\mu}{ds} \frac{dx^\nu}{ds},
\end{equation}
in \eref{F-rind} we have the force on the plates to be
\begin{equation}
F^3 = \left[ m_{\rm a} + m_{\rm b} \right]g.
\end{equation}
Thus, the total force on the Casimir apparatus, including 
the sum of the force on the vacuum energy, and the plates, is
\begin{equation}
F^3 = \left[ m_{\rm a} + m_{\rm b}
      + E_{\rm a}^{(0)} + E_{\rm b}^{(0)} + E_{\rm cas}^{(0)} \right]g.
\end{equation}
Interpreting $m_{\rm a,b}$ to be the bare masses of the plates
and renormalizing 
$m_{\rm a,b} + E_{\rm a,b}^{(0)} \rightarrow M_{\rm a,b}$
we thus conclude
\begin{equation}
F^3 = \left[ M_{\rm a} + M_{\rm b} + E_{\rm cas}^{(0)} \right]g,
\end{equation}
where $M_{a,b}$ are the renormalized masses of the plates.

%-------------------------------------------
\section{Rotating Casimir apparatus}
\label{rotation}

\begin{figure}
\begin{center}
\includegraphics{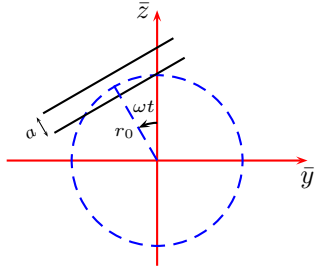}
\caption{\label{rot-c}
Casimir apparatus rotating with constant angular speed
$\omega$ about the $\bar{x}$ axis. The plane of the plates
are chosen to be perpendicular to the radial direction.
We have $r_a=r_0-\frac{a}{2}$, and $r_b=r_0+\frac{a}{2}$.}
\end{center}
\end{figure}

A scalar field $\bar{\phi}(\bar{x})$
in the presence of a background $\bar{V}(\bar{x})$
is described by the action
\begin{equation}
W = \int d^4\bar{x}
\left[
- \frac{1}{2} \eta^{ab} 
\bar{\partial}_a \bar{\phi}(\bar{x}) \bar{\partial}_b \bar{\phi}(\bar{x})
- \frac{1}{2} \bar{V}(\bar{x}) \bar{\phi}(\bar{x})^2
\right].
\label{W-bar}
\end{equation}
A Casimir apparatus, built out of so-called semi-transparent plates,
rotating with constant angular speed $\omega$ about the $\bar{x}$ axis,
as illustrated in \fref{rot-c}, will be described by the potential
\numparts
\begin{eqnarray}
\fl
\bar{V}(\bar{x}) 
&=& \lambda_a\,\delta (\bar{z}\cos\omega\bar{t}-\bar{y}\sin\omega\bar{t} - r_a)
+\lambda_b\,\delta (\bar{z}\cos\omega\bar{t} - \bar{y}\sin\omega\bar{t} - r_b)
\\
\fl
&=& \lambda_a\,\delta \left(\bar{z}\cos\omega\bar{t}
      - \bar{y}\sin\omega\bar{t} - r_0 + \frac{a}{2} \right)
+\lambda_b\,\delta \left(\bar{z}\cos\omega\bar{t} 
      - \bar{y}\sin\omega\bar{t} - r_0 - \frac{a}{2} \right).
\label{V-bar}
\end{eqnarray} 
\endnumparts
Making the following transformations
\numparts
\begin{eqnarray}
y &= +\bar{y}\cos\omega\bar{t}+\bar{z}\sin\omega\bar{t}, 
   \qquad\qquad &x=\bar{x},
\label{trans-a}
\\
z &= -\bar{y}\sin\omega\bar{t}+\bar{z}\cos\omega\bar{t} - r_0, 
\qquad \qquad &t=\bar{t},
\label{trans-b}
\end{eqnarray}
\endnumparts
which has the following inverse transformations
\numparts
\begin{eqnarray}
\bar{y} &= y\cos\omega t - (z+r_0)\sin\omega t, 
  \qquad\qquad &\bar{x}=x,
\label{in-trans-a}
\\
\bar{z} &= y\sin\omega t + (z+r_0)\cos\omega t, 
  \qquad \qquad &\bar{t}=t,
\label{in-trans-b}
\end{eqnarray}
\endnumparts
the action in \eref{W-bar} takes the form
\begin{equation}
W = \int d^4x \sqrt{-g(x)} \, {\cal{L}}(\phi(x)),
\end{equation}
where
\begin{equation}
{\cal{L}}(\phi(x)) 
= - \frac{1}{2} g^{\mu\nu}(x) \partial_\mu \phi(x) \partial_\nu \phi(x)
- \frac{1}{2} V(x) \phi(x)^2, 
\label{L}
\end{equation}
in which the background potential, 
\begin{equation}
V(x) = \lambda_a\,\delta \left(z + \frac{a}{2}\right) 
       + \lambda_b\,\delta \left(z - \frac{a}{2} \right),
\label{pot}
\end{equation}
is now independent of $t$. The metric is
\begin{equation}
g_{\mu\nu}(x) = 
\left[ \begin{array}{cccc}
-(1-\omega^2r^2) &0& -\omega (z+r_0) & \omega y \\ 0&1&0&0 \\
-\omega (z+r_0) &0& 1 & 0 \\ \omega y &0& 0 & 1 
\end{array} \right],
\label{rot-met}
\end{equation}
and its inverse is
\begin{equation}
g^{\mu\nu}(x) =
\left[ \begin{array}{cccc}
-1 &0& -\omega (z+r_0) & \omega y \\ 0&1&0&0 \\
-\omega (z+r_0) &0& \{1 - \omega^2 (z+r_0)^2\} & \omega^2 y (z+r_0) \\ 
\omega y &0& \omega^2 y (z+r_0) & (1 - \omega^2 y^2)
\end{array} \right],
\end{equation}
where $r^2 = y^2 + (z+r_0)^2$, and $g(x)={\rm det}\,g_{\mu\nu}(x)=-1$.
For the sake of bookkeeping we note
the corresponding nonzero components of the connection symbols:
$\Gamma^2_{00}=-\omega^2 y, \Gamma^3_{00} = -\omega^2 z$,
and $ \Gamma^3_{02} = - \Gamma^2_{03} = \omega$.
The energy-momentum tensor, or the stress tensor,
in the curvilinear coordinates is
\begin{equation}
t_{\alpha \beta}(x)
= \partial_\alpha \phi(x) \partial_\beta \phi(x)
  + g_{\alpha \beta}(x) {\cal{L}}(\phi(x)).
\label{tmunu}
\end{equation}
At the one-loop level the Green's function $G(x,x^\prime)$
is related to the fields by the correspondence
$\langle T \phi(x) \phi(x^\prime) \rangle =- \rmi \, G(x,x^\prime)$.
%In our case the Green's function is the solution to
%\begin{equation}
%- \left[ \partial_\mu g^{\mu\nu}(x) \partial_\nu - V(z) \right] 
%G(x,x^\prime) = \delta^{(4)}(x-x^\prime).
%\label{Gf}
%\end{equation}

%------------------------------------------------
\subsection{Centripetal force}

In terms of the curvilinear coordinates 
the potential \eref{pot} is independent of time $t$
which renders the energy-momentum tensor independent of time. 
Using this in \eref{FA-ct} we calculate the force
on the vacuum energy associated with the Casimir apparatus to be 
\begin{equation}
\bar{F}^a(\bar{t}) = \int d^3x \, t^{\mu0}(x) 
\,\frac{\partial}{\partial t} \frac{\partial \bar{x}^a(x)}{\partial x^\mu}.
\label{Ftmunu}
\end{equation}
We shall be interested in the leading order contributions 
to the force in the parameter $\omega r_0$.
To this end we observe that only the zeroth order term in the
energy-momentum tensor contributes, 
which is diagonal and a function of $z$ alone.
Using this in \eref{Ftmunu} we can write the
centripetal force on the vacuum energy to be 
\begin{eqnarray}
\fl
\bar{\bf F}(\bar{t}) = 
- \omega^2 \,\hat{\bar{\bf r}}(\omega \bar{t})\,r_0 \int d^3x \,t_{00}^{(0)}(z)
- \omega^2 \,\hat{\bar{\bf r}}(\omega \bar{t}) \int d^3x \, z \,t_{00}^{(0)}(z)
+ \omega^2 {\cal O}(\omega r_0),
\label{Ft00-1}
\end{eqnarray}
where the unit vector is
$\hat{\bar{\bf r}}(\theta) = \cos\theta\,\hat{\bar{\bf z}}
    -\sin\theta\,\hat{\bar{\bf y}}$.
We define the total energy associated with the field and 
the center of inertia (energy) as
measured in the curvilinear coordinates as 
\begin{equation}
E^{(0)}_{\rm tot} = \int d^3x \, t_{00}^{(0)}(x)
\qquad {\rm and} \qquad
z_{\rm cm} = \frac{1}{E^{(0)}_{\rm tot}} \int d^3x \,z \, t_{00}^{(0)}(x)
\label{E-Q}
\end{equation}
respectively, 
in terms of which the centripetal force in \eref{Ft00-1} can be 
written in the form
\begin{equation}
\bar{\bf F}(\bar{t}) = 
- \omega^2 \,\left[ r_0 + z_{\rm cm} \right] 
  \,\hat{\bar{\bf r}}(\omega \bar{t}) \, E_{\rm tot}^{(0)} 
= - \omega^2 \, \bar{r}_{\rm cm} 
  \,\hat{\bar{\bf r}}(\omega \bar{t}) \, E_{\rm tot}^{(0)}, 
\label{Ft00}
\end{equation}
where $\bar{r}_{\rm cm} = r_0 + z_{\rm cm}$, 
is the center of inertia in the local Lorentz coordinates.
$t_{00}^{(0)}(z)$ is the energy density of the %parallel 
plates when the %centripetal 
acceleration is switched off. In particular we recall 
\begin{eqnarray}
t_{00}^{(0)}(x) = \frac{1}{2\rmi} \left\{ \left( 
  \frac{\partial}{\partial t} \frac{\partial}{\partial t^\prime}
  - \frac{\partial^2}{\partial t^2} \right) G^{(0)}(x,x^\prime) 
  \right\}_{x=x^\prime} + \, \frac{1}{4\rmi} \, {\nabla}^2 G^{(0)}(x,x), 
\nonumber
\end{eqnarray}
where the Green's function is of the form
\begin{eqnarray}
G^{(0)}(x,x^\prime)
= \int_{-\infty}^{+\infty}
  \frac{d\omega}{2 \pi} \int \frac{d^2 k_\perp}{(2 \pi)^2}
  \, e^{-\rmi \omega (t - t^\prime)}
  e^{\rmi {\bf k}_\perp \cdot ({\bf x}_\perp - {\bf x}_\perp^\prime)}
  g^{(0)}(z,z^\prime;\kappa),
\nonumber
\end{eqnarray}
with $\kappa^2 = {\bf k}_\perp^2 - \omega^2$, 
and $g^{(0)}(z,z^\prime;\kappa)$ satisfies
\begin{equation}
- \left[ \frac{\partial^2}{\partial z^2} - \kappa^2 - V^{(0)}(z)
\right] g^{(0)}(z,z^\prime;\kappa) = \delta(z-z^\prime).
\label{zgf0}
\end{equation}
Using these relations the total energy per unit area 
in \eref{E-Q} takes the form
\begin{equation}
\frac{E^{(0)}_{\rm tot}}{A}
= - \frac{1}{2} \int_{-\infty}^{+\infty}
\frac{d\zeta}{2 \pi} \int \frac{d^2 k_\perp}{(2 \pi)^2}
\,2 \zeta^2 \int_{-\infty}^{+\infty} dz \, g^{(0)}(z,z;\kappa),
\label{Eg0}
\end{equation}
where we have switched to imaginary frequencies, 
$\omega \rightarrow \rmi \zeta$, thus 
$\kappa^2 = {\bf k}_\perp^2 + \zeta^2$.
Center of inertia in \eref{E-Q} can be similarly expressed 
in terms of the first moment of $g^{(0)}(z,z;\kappa)$.

%--------------------------------------------
\subsection{Centripetal force on a single plate}

A single plate rotating with constant angular speed is
described by a single delta function in the potential in \eref{pot}
(letting $a=0$, $r_0=r_a$, and $\lambda_b=0$),
\begin{equation}
V(z) = \lambda_a \, \delta (z),
\label{1-v0}
\end{equation}
and the corresponding solution to eq. (\ref{zgf0}) evaluated
at $z=z^\prime$ is 
\begin{equation}
g^{(0)}(z,z;\kappa) =
\frac{1}{2 \kappa} - \frac{\lambda_a}{\lambda_a + 2 \kappa}
  \frac{1}{2 \kappa} \, e^{-2\kappa|z|}
\label{1-g0-zzp}
\end{equation}
in terms of which the force in \eref{Ft00} evaluates to
\begin{equation}
\bar{\bf F}(\bar{t}) = - \hat{\bar{\bf r}}(\omega\bar{t}) 
       \, \omega^2 \bar{r}_{\rm cm} \, E_a^{(0)},
\end{equation}
where $\bar{r}_{\rm cm}=r_0=r_a$, and
\begin{equation}
\frac{E_{a}^{(0)}}{A}
= \frac{1}{12\pi^2} \int_0^\infty \kappa^2 d\kappa
\, \frac{\lambda_a}{\lambda_a + 2\kappa}
= \frac{1}{96\pi^2}
    \int_0^\infty \frac{dy}{y} \, \frac{y^3}{1+\frac{y}{\lambda_a}},
\label{Eva}
\end{equation}
a divergent quantity, is the energy per unit area associated
with a single plate. The center of inertia, $z_{\rm cm}$,
in \eref{E-Q} evaluates to zero for a single plate.

%-------------------------------------------
\subsection{Parallel plates}

Two parallel plates separated by a distance $a$ and rotating
with constant angular speed $\omega$ is described by the potential 
in \eref{pot}. The reduced Green's function, 
$g^{(0)}(z,z^\prime;\kappa)$, which is solution to \eref{zgf0}, 
evaluated at $z=z^\prime$, in the region $z<-\frac{a}{2}$ is 
\begin{eqnarray}
\fl
\frac{1}{2\kappa} - \frac{1}{2\kappa}\frac{e^{2\kappa z}}{\Delta}
\left[
\frac{\lambda_a}{2\kappa}
\left( 1 + \frac{\lambda_b}{2\kappa} \right) e^{\kappa a}
+ \frac{\lambda_b}{2\kappa}
\left( 1 - \frac{\lambda_a}{2\kappa} \right) e^{-\kappa a}
\right],
\end{eqnarray}
in the region $-\frac{a}{2}<z<\frac{a}{2}$ is 
\begin{eqnarray}
\fl
\frac{1}{2\kappa} - \frac{1}{2\kappa} \frac{e^{-\kappa a}}{\Delta}
\left[
\frac{\lambda_a}{2\kappa}
\left( 1 + \frac{\lambda_b}{2\kappa} \right) e^{-2\kappa z}
+ \frac{\lambda_b}{2\kappa}
\left( 1 + \frac{\lambda_a}{2\kappa} \right) e^{2\kappa z}
- 2 \frac{\lambda_a}{2\kappa} \frac{\lambda_b}{2\kappa}\, e^{-\kappa a}
\right],
\end{eqnarray}
and in the region $\frac{a}{2}<z$ is
\begin{eqnarray}
\fl
\frac{1}{2\kappa} - \frac{1}{2\kappa} \frac{e^{-2\kappa z}}{\Delta}
\left[
\frac{\lambda_a}{2\kappa}
\left( 1 - \frac{\lambda_b}{2\kappa} \right) e^{-\kappa a}
+ \frac{\lambda_b}{2\kappa}
\left( 1 + \frac{\lambda_a}{2\kappa} \right) e^{\kappa a}
\right],
\end{eqnarray}
where
\begin{equation}
\Delta = 
\left( 1 + \frac{\lambda_a}{2\kappa} \right) 
\left( 1 + \frac{\lambda_b}{2\kappa} \right) 
-\frac{\lambda_a}{2\kappa} \frac{\lambda_b}{2\kappa} \, e^{-2\kappa a}.
\end{equation}
Using the above expressions for $g^{(0)}(z,z;\kappa)$ in the 
appropriate regions of the $z$ integral in \eref{Eg0}
the centripetal force in \eref{Ft00} evaluates to
\begin{eqnarray}
\bar{\bf F}(\bar{t}) 
= - \hat{\bar{\bf r}}(\omega\bar{t}) \, \omega^2 \bar{r}_{\rm cm} 
\left[E_{\rm a}^{(0)} + E_{\rm b}^{(0)} + E_{\rm cas}^{(0)} \right],
\label{F-para}
\end{eqnarray}
where $E^{(0)}_{\rm a,b}$ are the total energies
due to single plates given in terms of \eref{Eva},
$E^{(0)}_{\rm cas}$ is the Casimir energy given as
\begin{equation}
\frac{E_{\rm cas}^{(0)}}{A}
= - \frac{1}{12 \pi^2} \int_0^\infty \kappa^2 \, d\kappa
\left[ 2 \kappa a
+ \frac{1}{1 + \frac{\lambda_a}{2\kappa}}
+ \frac{1}{1 + \frac{\lambda_b}{2\kappa}} \right]
\frac{1}{2\kappa} \frac{\partial}{\partial a} \ln \Delta,
\label{Ecas}
\end{equation}
and $\bar{r}_{\rm cm}=r_0 + z_{\rm cm}$,
where the center of inertia in the curvilinear coordinates, $z_{\rm cm}$, 
for parallel plates evaluates to
\begin{equation}
z_{\rm cm} \, E_{\rm tot}^{(0)} = 
- \frac{a}{24\pi^2} \int_0^\infty \kappa^2 d\kappa \frac{1}{\Delta}
\left[ \frac{\lambda_b}{2\kappa} - \frac{\lambda_a}{2\kappa} \right],
\label{zcm-para}
\end{equation}
which is in general not zero because of the asymmetry in the couplings,
but evaluates to zero when $\lambda_a = \lambda_b$,
or in the Dirichlet limit.

%-------------------------------------------
\section{Orientation independence of centripetal force}
\label{sec-or}

\begin{figure}
\begin{center}
\includegraphics{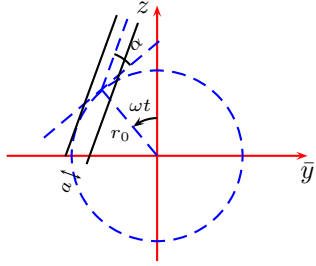}
\caption{\label{or-in-cir}
A Casimir apparatus oriented at an angle $\alpha$ with respect to 
the tangent to the circle of radius $r_0$ and rotating with 
constant angular speed $\omega$ about the $\bar{x}$ axis.
We have $r_a\cos\alpha = r_0\cos\alpha - \frac{a}{2}$,
$r_b\cos\alpha = r_0\cos\alpha + \frac{a}{2}$,
$r_0=(r_a+r_b)/2$, and $a=(r_b-r_a) \cos\alpha$. }
\end{center}
\end{figure}

A Casimir apparatus making an arbitrary angle $\alpha$ with respect to
the tangent while rotating with constant angular speed $\omega$,
as illustrated in \fref{or-in-cir}, is described by the potential, 
(displaying only for one plate to save typographic space,)
\numparts
\begin{eqnarray}
\bar{V}(\bar{x}) &=& \lambda_a\,\delta 
(\bar{z}\cos(\alpha+\omega\bar{t}) 
 - \bar{y}\sin(\alpha+\omega\bar{t}) - r_a \cos\alpha)
\\
&=& \lambda_a\,\delta
\left(\bar{z}\cos(\alpha+\omega\bar{t})
 - \bar{y}\sin(\alpha+\omega\bar{t}) - r_0 \cos\alpha + \frac{a}{2} \right).
\label{V-bar-or}
\end{eqnarray}
\endnumparts
We make the transformations
\numparts
\begin{eqnarray}
\fl
y &= +\bar{y}\cos(\alpha+\omega\bar{t})
     +\bar{z}\sin(\alpha+\omega\bar{t})
     - r_0 \sin\alpha, \qquad x&=\bar{x},
\\
\fl
z &= -\bar{y}\sin(\alpha+\omega\bar{t})
     +\bar{z}\cos(\alpha+\omega\bar{t})
     - r_0 \cos\alpha, \qquad t&= \bar{t},
\end{eqnarray}
\endnumparts
which has the inverse transformations
\numparts
\begin{eqnarray}
\fl
\bar{y} &= (y+r_0\sin\alpha) \cos(\alpha+\omega\bar{t})
           -(z+r_0\cos\alpha) \sin(\alpha+\omega\bar{t}),
           \qquad \bar{x}&=x,
\\
\fl
\bar{z} &= (y+r_0\sin\alpha) \sin(\alpha+\omega\bar{t})
           +(z+r_0\cos\alpha) \cos(\alpha+\omega\bar{t}),
           \qquad \bar{t}&=t,
\end{eqnarray}
\endnumparts
which reduces to the transformations in \eref{trans-a} and \eref{trans-b}
for $\alpha =0$.
The metric corresponding to this transformation 
evaluates to be
\begin{equation}
\fl
g_{\mu\nu}(x) =
\left[ \begin{array}{cccc}
-(1-\omega^2r^2) &0& -\omega (z+r_0\cos\alpha) & \omega (y+r_0\sin\alpha) \\
 0&1&0&0 \\
-\omega (z+r_0\cos\alpha) &0& 1 & 0 \\ 
+\omega (y+r_0\sin\alpha) &0& 0 & 1
\end{array} \right],
\end{equation}
where $r^2 = (y+r_0\sin\alpha)^2 + (z+r_0\cos\alpha)^2$,
and reduces to the metric in \eref{rot-met} for $\alpha =0$.
Thus on, repeating the steps in \sref{rotation},
we evaluate the centripetal force to be
\begin{equation}
\bar{\bf F}(\bar{t})
= - \omega^2 \, \bar{\bf r}_{\rm cm}(\omega\bar{t})
\left[E_{\rm a}^{(0)} + E_{\rm b}^{(0)} + E_{\rm cas}^{(0)} \right],
\label{0r-Ft00}
\end{equation}
where the center of inertia in the local Lorentz coordinates is
\begin{equation}
\bar{\bf r}_{\rm cm}(\omega \bar{t}) = 
r_0 \hat{\bar{\bf r}}(\omega \bar{t})
+ z_{\rm cm} \hat{\bar{\bf r}}(\alpha + \omega \bar{t}),
\end{equation}
in which the center of inertia in the curvilinear coordinates,
$z_{\rm cm}$, is again given using \eref{zcm-para}.
We observe that if the origin of the curvilinear coordinates is
chosen such that $z_{\rm cm}=0$ then the centripetal force
is independent of the orientation of the Casimir apparatus. 
We shall end by pointing to the discussion on frame dependence of
center of mass in page 176 of reference~\cite{moller} 
and to the original work cited in it.

%-------------------------------------------
%\section{Accelerating frames of reference}
%\subsection{Galilean metric}
%\subsection{Non relativistic acceleration}
%\subsection{Rindler metric}
%\section{Time dependent potential as a perturbation}
%\section{Orientation independence - Gauge independence}
%\subsection{Rindler metric}
%\subsection{Rotation metric}
%\section{Sphere}
%----------------------------------------------
\ack
We are grateful to Iver Brevik, In\'{e}s Cavero-Pel\'{a}ez, Simen Ellingsen,
Steve Fulling, Klaus Kirsten, August Romeo, Aram Saharian, and Justin Wilson 
for comments and helpful suggestions.
This work was supported in part by a Collaborative Research Grant from
the US National Science Foundation (PHY-0554926) and in part by the
US Department of Energy (DE-FG02-04ER41305).  We are grateful to Michael
Bordag for arranging such a fruitful QFEXT07 workshop.

%----------------------------------------------

\end{document}